\documentclass{article}

\usepackage{arxiv}
\usepackage[utf8]{inputenc} 
\usepackage[T1]{fontenc}    
\usepackage{hyperref}       
\usepackage{url}            
\usepackage{booktabs}       
\usepackage{amsfonts}       
\usepackage{nicefrac}       
\usepackage{microtype}      
\usepackage{graphicx}
\usepackage{float}          
\usepackage[justification=centering]{caption}
\usepackage{subcaption}      
\title{GreenBox: Prototyping of an Automatic Road Accident Detection System with Real-Time Notification SMS%
\thanks{Affiliations: $^{1}$INESC TEC --- INESC Technology and Science, 4200-465 Porto, Portugal.
$^{2}$Engineering Department, School of Science and Technology, Quinta de Prados, University of Trás-os-Montes and Alto Douro, 5000-911 Vila Real, Portugal.
$^{3}$IEETA --- Institute of Electronics and Informatics Engineering of Aveiro, Campus Universitário de Santiago, 3810-193 Aveiro, Portugal.
$^{4}$Altice Labs, 3810-106 Aveiro, Portugal.
$^{\dagger}$These authors contributed equally to this work.
$^{*}$Corresponding author: avalente@utad.pt}}

\author{
  Júlio Rocha$^{1,2,\dagger}$ \\
  INESC TEC -- INESC Technology and Science\\
  University of Trás-os-Montes and Alto Douro\\
  \texttt{al83013@alunos.utad.pt}\\
  \And
  António Valente$^{1,2,\dagger,*}$ \\
  INESC TEC -- INESC Technology and Science\\
  University of Trás-os-Montes and Alto Douro\\
  \texttt{avalente@utad.pt}\\
  \And
  Salviano Soares$^{2,3,\dagger}$ \\
  University of Trás-os-Montes and Alto Douro\\
  IEETA -- Institute of Electronics and Informatics Engineering of Aveiro\\
  \texttt{salblues@utad.pt}\\
  \And
  Filipe Cabral Pinto$^{4,\dagger}$ \\
  Altice Labs\\
  Aveiro, Portugal\\
  \texttt{filipe-c-pinto@alticelabs.com}\\
}

\begin{document}
\maketitle

\begin{abstract}
The Internet of Things (IoT) project, called "GreenBox", proposes the development of a prototype for the detection of road accidents, using sensors and actuators connected to an Arduino Uno \cite{uno} and a Global System for Mobile Communications (GSM) card with 4G support, SIM7600G-H (global version) from DF-Robot\cite{sim7600v1}. This system sends Short Message Service (SMS) to pre-established contacts, alerting, for example, family or friends, so that they immediately contact emergency entities in the event of an accident and provide them with all the necessary information, such as the location of the vehicle. The sensors include four push-buttons, with a resistance of $10k\Omega$, in order to define their default logical state \cite{McRoberts2013}, representing impact or impact sensors for each side of the vehicle; two water level sensors for the engine compartment and trunk; and a gyroscope/accelerometer to detect rollover from a 70 degrees inclination. The prototype also has a relay that is activated to turn off the engine in the event of a detected accident, preventing further damage. The GSM board has a Global Positioning System (GPS) antenna attached, allowing it to locate the vehicle and determine its speed, moments before the possible accident. The system also allows you to turn off the vehicle via SMS in case of theft. The entire project is prepared for the owner, for example the driver of the vehicle, to cancel all automation, via SMS, in case of false accident detection. Rollover detection is calculated using the arctangent of the accelerometer values and instructions for sending notifications are carried out by AT (Attention Commands) \cite{atcommander} \cite{1oT} commands, between the microcontroller and the SIM7600 shield.
\end{abstract}


\section{Introduction}

\begin{figure}[H]
\centering
\includegraphics[width=13.0 cm]{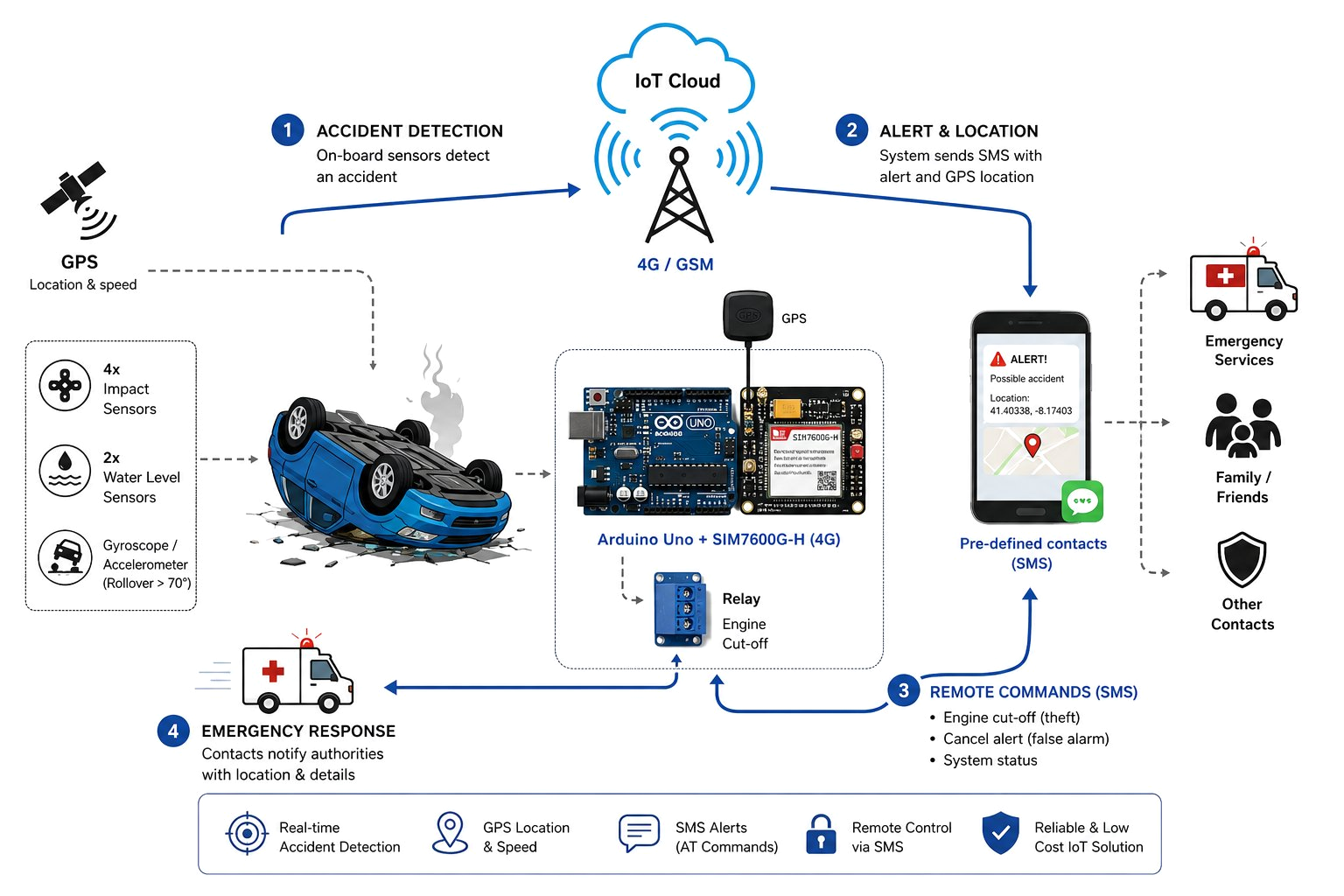}
\caption{GreenBox: Graphical Abstract\label{fig1}. After the accident is detected: (1) The system sends location and speed to the network provider, (2) the network provider interprets this information and forwards it to the recipient specified in the SMS sent by the system, (3) the recipient then alerts emergency authorities, providing them with the location, and (4) emergency services proceed to the accident local.}
\end{figure}

\begin{figure}[H]
\centering
\includegraphics[width=10.0 cm]{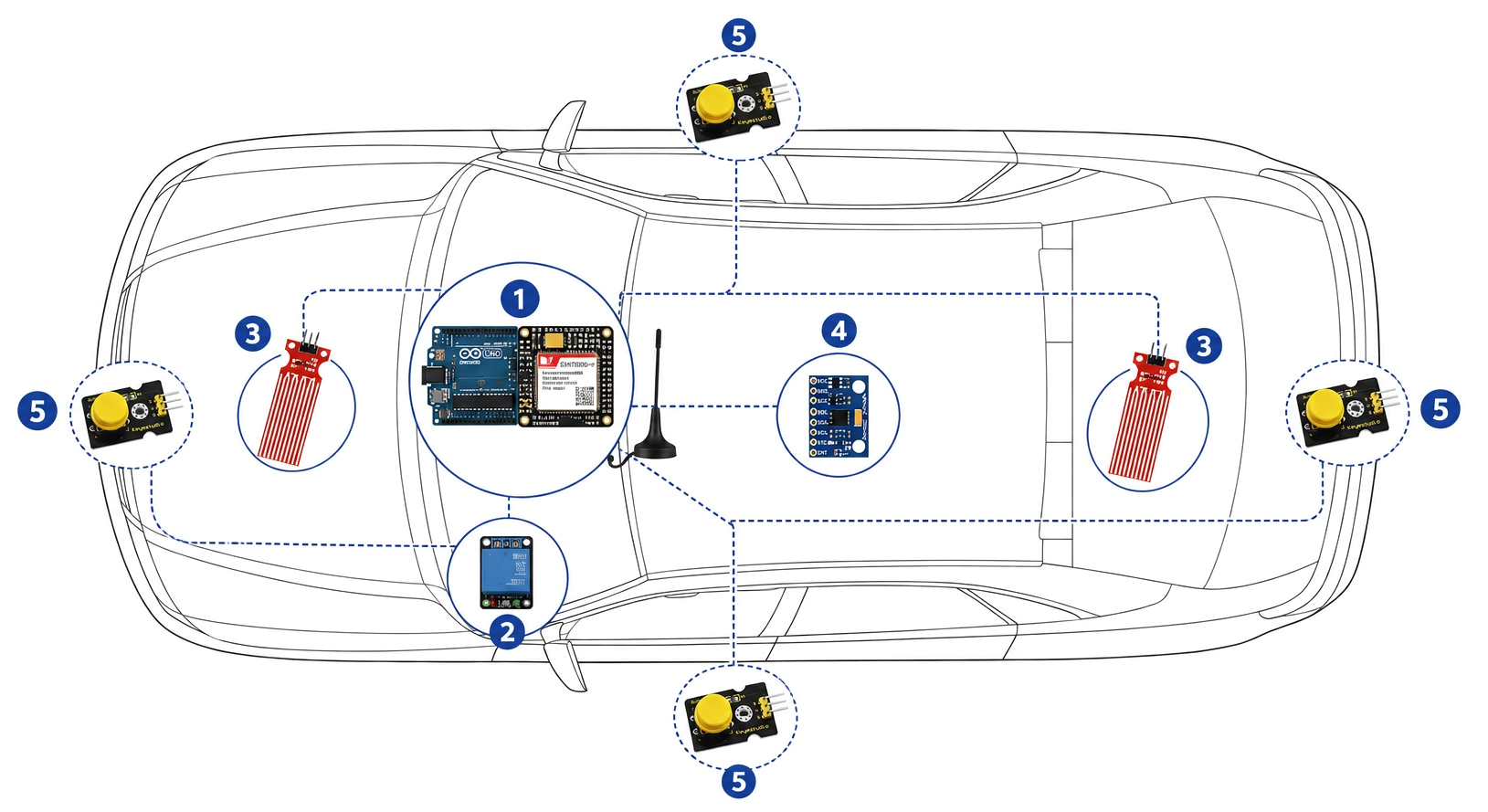}
\caption{GreenBox: Proposed hardware distribution schema\label{fig2}. (1) Arduino and Sim7600 shield, (2) relay actuator, (3) water level sensor, (4) gyroscope and accelerometer sensor, and (5) push-buttons.}
\end{figure}

In recent years, the increase in road accidents has generated significant concerns in terms of safety and efficiency in responding to emergencies \cite{simones2015impacto} \cite{silva2016acidentes} \cite{alonso2020sig}. This project, as shown in Figure~\ref{fig1}, addresses these concerns by developing an innovative accident detection system that uses accessible and effective technologies. Using a combination of sensors and actuators, connected to a microcontroller, and a communication module, strategically installed in a car, as represented in Figure~\ref{fig2}, the system not only detects a possible road accident, but also quickly communicates the location and situation of the vehicle to one or more defined contacts, and consequently to the competent authorities. The use of Arduino as a central platform allows for effective integration of several sensors, including push-buttons \cite{pushbutton} for crash detection, as well as a gyroscope/accelerometer, model MPU-6050 \cite{gyroaccel}, to monitor dangerous and sudden inclines. Also the inclusion of water level sensors \cite{water} provides additional protection against flood damage. In addition, the system is designed to automatically shut down the engine in the event of an accident, minimizing additional risks. The system's ability to send SMS notifications using the SIM7600 board is crucial for rapid communication with emergency services, if so programmed. GPS integration allows for a very close location of the vehicle, facilitating a more effective urgent response. This project not only contributes to road safety, and can save lives, but also offers additional features, such as the possibility of turning off the engine remotely in case of theft, increasing the vehicle's security. Based on these functionalities, the project proposes, not only to reduce the response time in emergencies, but also to provide a practical and scalable solution for road safety. The rest of the article is organised as follows. Section 2 presents the related work. Section 3 describes the proposed structure. The experimental work is presented in section 4. Section 5 deals with the results observed during the tests and finally the conclusions are presented in Section 6.

\section{Related Work}

The research of road accident detection and emergency communication systems has been a topic of growing interest in the scientific community, given its importance for public safety and efficiency in responding to emergencies. This review addresses several studies that provide valuable insights for the development of the "GreenBox" project.

A fundamental study in this area is the "IoT Based Automatic Vehicle Accident Detection And Rescue System", which explores the use of an accelerometer to monitor the variation in the speed of a vehicle in the event of a frontal or rear-end collision, a sensor used in the project for rollover detection. However, this study may be insufficient for the effective detection of the accident, due to the probability of false positives, in case of sudden braking, as well as in sending the location that does not yet exist, but is foreseen in the chapter on future work. In addition, communication is carried out via HTTP protocol \cite{http}, through a ESP8266 wifi card, embedded in the Node MCU card, paired with a hotspot present in the vehicle, which can be the driver's smartphone \cite{Dashora2020}.

The "Automatic Vehicle Accident Detection and Messaging System" project presents an approach similar to the "GreenBox", proposing a system that uses a microcontroller embedded in an Arduino Uno board, a GSM module for sending SMS as a notification, a vibration sensor (SW-420), an accelerometer and a Liquid Crystal Display (LCD) for auxiliary information. This work emphasises efficiency in data transmission and emergency response, crucial aspects that are also addressed in the project proposed in this article, through the use of sensors connected to the Arduino and fast communication via SMS \cite{KARTHIK20233124}.

Another study, entitled "Intelligent vehicle black box using IoT" discusses a monitoring and warning system based on an intelligent "black box", capable of detecting excessive alcohol consumption, recording acceleration data and sending alerts in case of accidents, in addition to tracking the vehicle's location. An interesting feature of this project is the detection of alcohol carried out by an analogue sensor, specifically the MQ3 module, which detects the presence of alcohol gases inside the vehicle, in concentrations ranging from 0.05 mg/L to 10 mg/L. The sensor measures conductivity, which increases with the concentration of alcohol, allowing the system to identify when the alcohol level of the driver or other occupant exceeds the safe limit \cite{AnilKumar2018}.

The research "Accident Prevention and Safety Assistance Using IoT and machine learning (ML)" presents an innovative system that uses the IoT and ML to monitor driver safety, focusing on the detection of fatigue, alcohol consumption and toxic gases. Equipped with Raspberry Pi, Pi camera and sensors, the device collects data in real time and classifies drivers as safe or risky, sending SMS alerts and storing information in the cloud. The main objective is to reduce road accidents, especially those caused by fatigue and driving under the influence of alcohol. Although the project does not use Machine Learning, this study reinforces the relevance of fast data collection and processing for road safety \cite{Uma2022}.

The work "IoT Based Car Accident Detection and Notification Algorithm for General Road Accidents" proposes a system for detecting and reporting road accidents based on the IoT, using Arduino, accelerometer sensor, vibration sensor, heart rate sensor built into the seat belt, as well as GPS and GSM modules for location and SMS communication. The aim of the study is to develop an algorithm that detects accidents in real time and immediately notifies emergency services, aiming to reduce response time and save lives, while monitoring the health of vehicle occupants \cite{IJECE16079}.

In addition to these studies, it is important to highlight the Advanced Driver Assistance Systems (ADAS), which have been widely investigated as a way to reduce human errors in driving and minimize road accidents. ADAS are defined as a set of technologies that improve the safety and efficiency of road traffic, offering assistance to the driver in various situations. These systems use sensors and communication between vehicles to prevent accidents and make driving easier. With the aim of reducing reliance on human skills and increasing safety on the roads, these systems use sensors to assist in driving and include features such as automatic speed control and collision warning. This study \cite{bengler2014} analyses the evolution of ADAS since the 1980s, highlighting the slow transition from research to production and the recent significant market penetration. It emphasizes the importance of appropriate assessment methods to ensure the safety of assistance functions, especially in complex scenarios, and the need to adapt systems to the needs of specific groups of drivers. In addition, it discusses the potential social and economic impact of automation and the importance of interdisciplinary research to facilitate the introduction of fully automated driving technologies.

Additionally, a study revealed that vehicles equipped with ADAS showed a significant reduction in accident rates on the road, involving the most vulnerable drivers \cite{Isaksson_Hellman_Lindman_2023}. Other research reflects the impact of ADAS systems on reducing road accidents, suggesting that the adoption of these technologies could prevent a considerable percentage of vehicle accidents \cite{Aleksa2024}.

Today, ADAS are constantly evolving, incorporating technologies such as cameras, radar sensors, and vehicle-to-vehicle communication to improve road safety, especially in interactions with cyclists. Although these systems have been shown to be effective in reducing accidents by improving risk recognition and reducing drivers' reaction times, their effectiveness can be compromised by several factors such as adverse weather conditions, complexity of the road environment, driver behaviour and sensor quality. Thus, it is essential to address these challenges to maximize the potential of ADAS in promoting safer driving \cite{Useche2024}.

These studies provide a solid basis for the development of the road accident detection system with SMS communication. They show the technical feasibility of the project's core components, such as the use of sensors to detect collisions and dangerous slopes, the integration of GPS modules for precise location and speed recording of the vehicle, and the use of SMS to send alerts quickly. In addition, they underline the importance and potential impact that these systems can have in reducing the severity of road accidents and improving the effectiveness of emergency responses.

The "GreenBox" project stands out for its comprehensive approach, combining multiple impact sensors (push-buttons) for accurate collision detection; water level sensors to alert when floods are detected; automatic engine shutdown system to prevent further damage; Remote vehicle shutdown functionality in case of theft.

These features, together with SMS communication and GPS location, elevate the "GreenBox" as a robust, practical and versatile solution for road safety.

\section{Proposed Framework}

This work proposes a framework for the detection of road accidents, using a combination of sensors and actuators, connected to an ATmega328 microcontroller \cite{atmega328p} attached to an Arduino board. In turn, the Arduino board is connected to a DF-Robot SIM7600 board, responsible for sending the programmed alerts, automatically, to one or more predefined contacts, when an accident is detected.

\begin{figure}[H]
\centering
\includegraphics[width=12.0 cm]{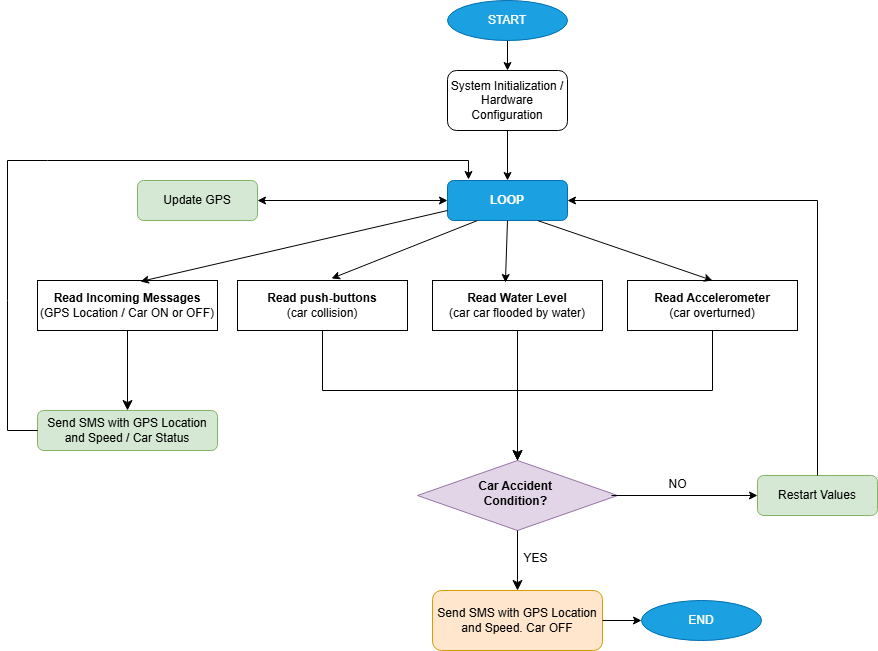}
\caption{GreenBox: Data Flow Diagram System\label{fig3}.}
\end{figure}

The data flow diagram of the IoT solution that supports the framework is shown in Figure~\ref{fig3}.

In the proposed framework, the following hardware will be used for proof of concept to monitoring the vehicle: an Arduino Uno board, composed of an ATmega328 microcontroller; a board composed of a chip (sim7600) responsible for SMS communication; four push-buttons that represent the impact/impact sensors; a gyroscope/accelerometer sensor responsible for rollover detection; two water level sensors, which will detect water in some compartment of the vehicle, if it does not overturn or suffer a violent impact, but crashes and enters, for example, a river; a relay actuator to disconnect that will be connected, for example, to the central unit of the motor, whose function will be to turn off the motor, at the time of the detected accident, to avoid further damage; an LCD screen to monitor the status of the system, a breadboard to support the connection tests between sensors, actuators and boards and, finally, the connection to the car battery through a converter (cigarette lighter charger) responsible for converting the voltage of the car battery, in this case 12V, to the voltage needed (5V) to power the system, in direct current (DC). The Arduino Uno as a microcontroller board based on the ATmega328. It includes 14 digital input/output pins (6 of which can be used as PWM outputs), 6 analog inputs, a USB connection, a power connector, and a reset button. The Arduino Uno is an open source product that integrates a processor and a bootloader, ready to be connected via USB to the computer.

The DF-Robot board with the SIM7600 chip, allows the microcontroller to be connected to a telecommunications network. The communication of the SIM7600 chip with the Arduino is carried out via series, using the RX and TX pins, and the instructions for real-time notifications, through the sending of messages, are initiated by the microcontroller and processed by this chip. The board SIM7600 used in this project has a module that offers 4G LTE connectivity, ideal for IoT projects that need fast Internet. In addition to allowing high-speed data transmission, it also supports location systems such as GPS, which is perfect for projects that require the exact location of objects, such as a vehicle in the event of an accident. It also supports several protocols, such as Transmission Control Protocol/Internet Protocol (TCP/IP), Hypertext Transfer Protocol (HTTP), Hypertext Transfer Protocol Secure (HTTPS) and Domain Name System (DNS), allowing complex network communications using AT commands. Through the various AT commands, it is possible to make calls and send messages, offering a robust and reliable solution for communication and automation.

In short, the "GreenBox" project stands out for its comprehensive approach, combining multiple impact sensors (push-buttons) for accurate collision or collision detection; water level sensors for protection and consequently warning against floods; automatic engine shutdown system to prevent further damage; Remote vehicle shutdown functionality in case of theft.

These features, together with communication via SMS and GPS location, differentiate the "GreenBox" as a practical, robust and versatile solution for road safety.

\section{Experimental Work}

During this section, the main solutions will be described, mainly in Arduino, to fulfill the objective of the proposed project, specifically in the use of AT commands, for the various instructions processed by the SIM7600 chip; the exploitation of the GPS module; the importance of the gyroscope/accelerometer and sending notifications via SMS. Other features implemented in the project, such as the use of push-buttons and water level sensors, will be presented superficially.

\begin{figure}[H]
\centering
\includegraphics[width=12.0 cm]{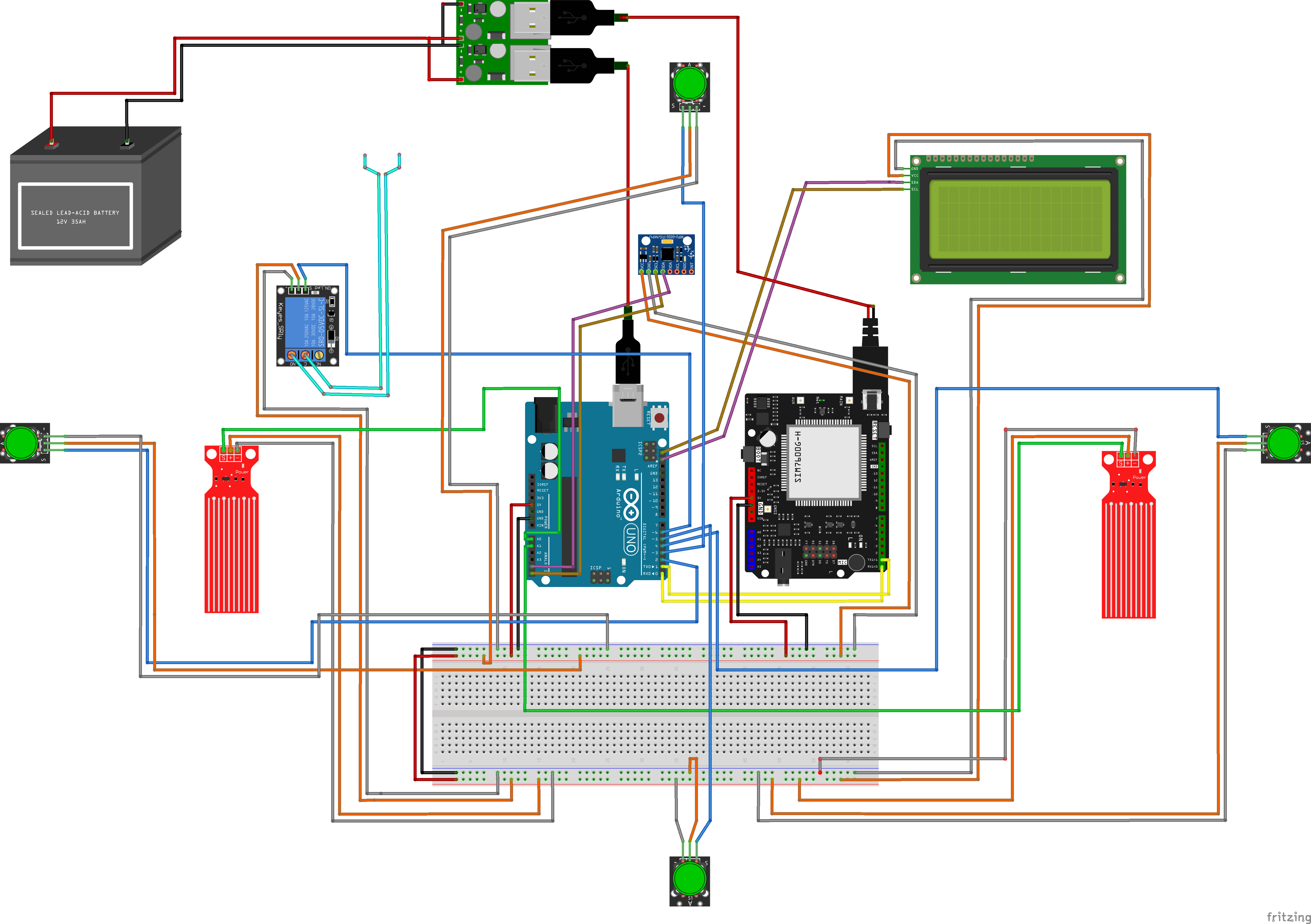}
\caption{GreenBox: Hardware Connection Scheme\label{fig4}.}
\end{figure}

After assembling the hardware, represented in Figure~\ref{fig4}, which includes the integration of sensors and actuators with the Arduino Uno and the SIM7600G-H board, the system programming was carried out in order to ensure the necessary functionalities.

The Arduino code, developed in the C programming language \cite{c}, was developed to manage sensor data, generate the necessary information and send SMS notifications in real time.

Initially, the main concern was to ensure that the system was fully functional during its startup. A critical aspect in the system is the start-up process, particularly when it comes to reading AT commands. The following subsections describe the main steps for configuring the sensors and actuators connected to the Arduino board, as well as the purpose of their use in the "GreenBox" project.

\subsection{Loading AT Controllers}

On the microcontroller it is implemented with a robust mechanism to ensure that all required AT commands are successfully loaded before the system becomes fully operational.

The system attempts to boot until it receives an "OK" response to the basic AT command, ensuring that the SIM7600 module responds correctly before proceeding. If the "OK" response is not received, the system waits 10 seconds before trying again. After the initial confirmation, the system sequentially loads a series of crucial AT commands, such as registration in the telecommunications network (AT+CREG? and AT+CGREG?); SMS configuration (AT+CMGF=1); definition of messaging service parameters (AT+CSMP=17,167,2,0), selection of GSM character set (AT+CSCS="GSM"); automatic activation of GPS (AT+CGPSAUTO=1); configuration of the notification mode for new messages (AT+CNMI=2,1,0,0,0) and cleaning of all messages stored on the SIM card (AT+CMGD=,4), in order to ensure space for new messages. By default, the Subscriber Identity Module (SIM) card used during the tests supports a total of 30 messages in memory. After successful configuration, the system sends a confirmation SMS ("SYSTEM STARTED!") to a predefined number. The selections of possible commands for the development of this project were studied using the AT Commander Tester software \cite{atcommander2}.

\subsection{GPS Module Exploration (Vehicle Location)}

One of the particularities of the GPS module, connected to the SIM7600 board, is the extraction of coordinates in the format of degrees and minutes, as indicated in the official documentation \cite{sim7600v1}.

For a better understanding of the extracted coordinates, they were converted to decimal degrees, making it easier to send and interpret this information in digital mapping systems, such as Google Maps, used during road tests, which we can see in the next section.

In addition to the coordinates, which guarantee the position of the vehicle, the speed provided by the GPS, originally in Knots, was converted to km/h (kilometres per hour), using the ratio of 1 Knot = 1,852 km/h. This allowed for a more intuitive understanding of the car's speed.

\subsection{Gyroscope/Accelerometer Exploration (Rollover Accident Detection)}

In addition to the GPS feature, another component, connected to the Arduino board, is the gyroscope/accelerometer, as this is essential for the accurate detection of accidents, specifically rollover. when the function to check the accelerometer values, plays a crucial role in this process, as it continuously monitors (\textasciitilde3 every 3 seconds) the vehicle's orientation in terms of Pitch (forward and backward inclination) and Roll (lateral inclination). The aforementioned function is responsible for reading the raw data from the accelerometer and calculating those angles, which are essential for rollover detection.

To summarise, the function uses the Wire library to communicate with the MPU-6050 sensor, in order to obtain the raw values of the accelerometer's X, Y, and Z axes, which are converted to acceleration units (G's) by dividing by 16384.0, which is the sensor's default sensitivity.

The calculation of Pitch and Roll \cite{pitchroll} is possible through the arctangent function (atan2), which returns the values in radians and later in degrees, by multiplying by 180 and consequently dividing by PI (3.14) \cite{Lindkvist2019}. These calculations are key to determining if the vehicle is on a dangerous incline. If the Pitch or Roll exceeds a limit (70 degrees) defined in the function that detects the accident, the system interprets the reading as a possible rollover, triggering emergency protocols, such as sending SMS and engine shutdown.

\subsection{Water Level Sensors (immersion accident detection)}

In addition to accident detection, through the data obtained by the accelerometer, the water level sensors are also integrated components in the "GreenBox" system, offering an additional layer of safety. Strategically installed in the engine compartment and trunk, these sensors play a vital role in identifying potentially dangerous situations that may go unnoticed by other detection mechanisms.

The main function is to detect the presence of water at abnormal levels, which may indicate that the vehicle has entered a river or a flooded area, jeopardizing the safety of the vehicle's occupants. This feature is particularly important in scenarios where the vehicle swerves without necessarily flipping over or colliding with an object, but ends up being in an equally dangerous situation.

When the sensors detect a water level above the set limit, the system interprets this as an accident situation. This detection immediately triggers the emergency function of the "GreenBox", which includes automatic engine shutdown to prevent further damage and SMS notifications with the vehicle's location. This approach ensures that even in road accidents with a lower probability, such as partial or total submersion of the vehicle, compared to a collision or rollover, the system is able to respond promptly, and can save lives in situations where every second is crucial.

\subsection{Push-Buttons (crash detection)}

The "GreenBox" system acts whenever an accident is detected, whether through rollover, abnormal water detection or collision. Push buttons were used to detect collisions, strategically positioned on each side of the vehicle. These buttons were not simply installed and programmed; The implementation involved crucial technical considerations to ensure reliability and accuracy. Each push-button module contains a resistance of $10k\Omega$, essential for establishing a stable logic state when the button is not pressed, avoiding fluctuating readings that could result in false positives. In addition, a "debounce" logic was implemented in the software, using a 100-millisecond delay to filter out rapid oscillations in the button signal, common in mechanical switches. This "debounce" technique prevents multiple accidental detections from a single press (button pressed). When any of these buttons is pressed, indicating a possible impact or collision, the "GreenBox" system responds promptly, turning off the vehicle's engine and sending an SMS notification with the precise location of the vehicle, similar to what is expected with the other sensors. The push-buttons were chosen for collision detection due to their simplicity, reliability, and robustness against false positives. Unlike inertial sensors such as the MPU-6050 accelerometer, which are sensitive to vibrations or and abrupt movements, mechanical buttons only trigger in the event of direct physical contact, eliminating the need for complex filtering and reducing the rate of false alarms. Additionally, they are significantly more cost-effective and straightforward to implement, without requiring constant calibration or advanced processing.

\subsection{Sending SMS from the system}

As already mentioned, one of the potentialities of the "GreenBox" project is the sending of notifications, via SMS, in real time. This feature is one of the crucial parts of the system, responsible for sending detailed notifications, via SMS, when the accident is detected.

The function starts by creating a "message" string with detailed information about the accident. This string includes some pertinent information for rapid assistance in the event of an emergency, such as latitude and longitude accurate to 6 decimal places, which translates to the vehicle's last known exact location. The last known speed of the vehicle is also attached and sent, moments before the possible accident detected. The status of the relay (which controls the car's engine) is also checked and informs if the car is on or off. It is planned that the vehicle's engine is turned off, as it is one of the functions of the relay at the time of the detected accident. This information will only be used to control the system. The function is completed by including a link to Google Maps, with the exact coordinates for easy and quick location.

The sending of the message uses the AT command "AT+CMGS", where the destination number is indicated, to start the process of sending SMS. After a short delay, the full message is sent using the sim7600.print(message) method. The ASCII character "26" (CTRL+Z) is sent to finish the process and send the SMS.

The variable "sim7600" specifically represents an object of the SoftwareSerial class, initially declared in the Arduino code. This declaration creates a SoftwareSerial instance named "sim7600", which is used for RX (receive) and TX (transmit) serial communication with the SIM7600 module.

Throughout the code, the "sim7600" is used to send AT commands and receive responses from the GSM module, allowing operations such as sending SMS or checking GPS status.

\subsection{Receiving SMS in the system}

Another particularity of the "GreenBox" system, in addition to sending automatic notifications when a (possible) accident is detected, is the reception and interpretation of messages sent by the owner of the system. During the tests, four predefined messages were programmed that will make the system act as desired: Know where the vehicle is; cancel the automatic mechanism when a road accident is falsely detected; switch off the motor and allow it to be started again if it is switched off, i.e. with the relay open.

The function begins by checking for new data in the buffer of the SIM7600 module, looking for "+CMTI:" in the response, which indicates a new message. If a new message exists, the index is extracted. Then, the AT command "AT+CMGR" is used to read the message, where the contact (number) of the sender and the content of the message are then extracted. The content of the message will be checked and if messages such as "status", "no", "lock" and "unlock" are found, the system will act as planned. After each message has been processed, it is deleted from the SIM card.

\section{Results}

\begin{figure}[H]
\centering
\includegraphics[width=12.0 cm]{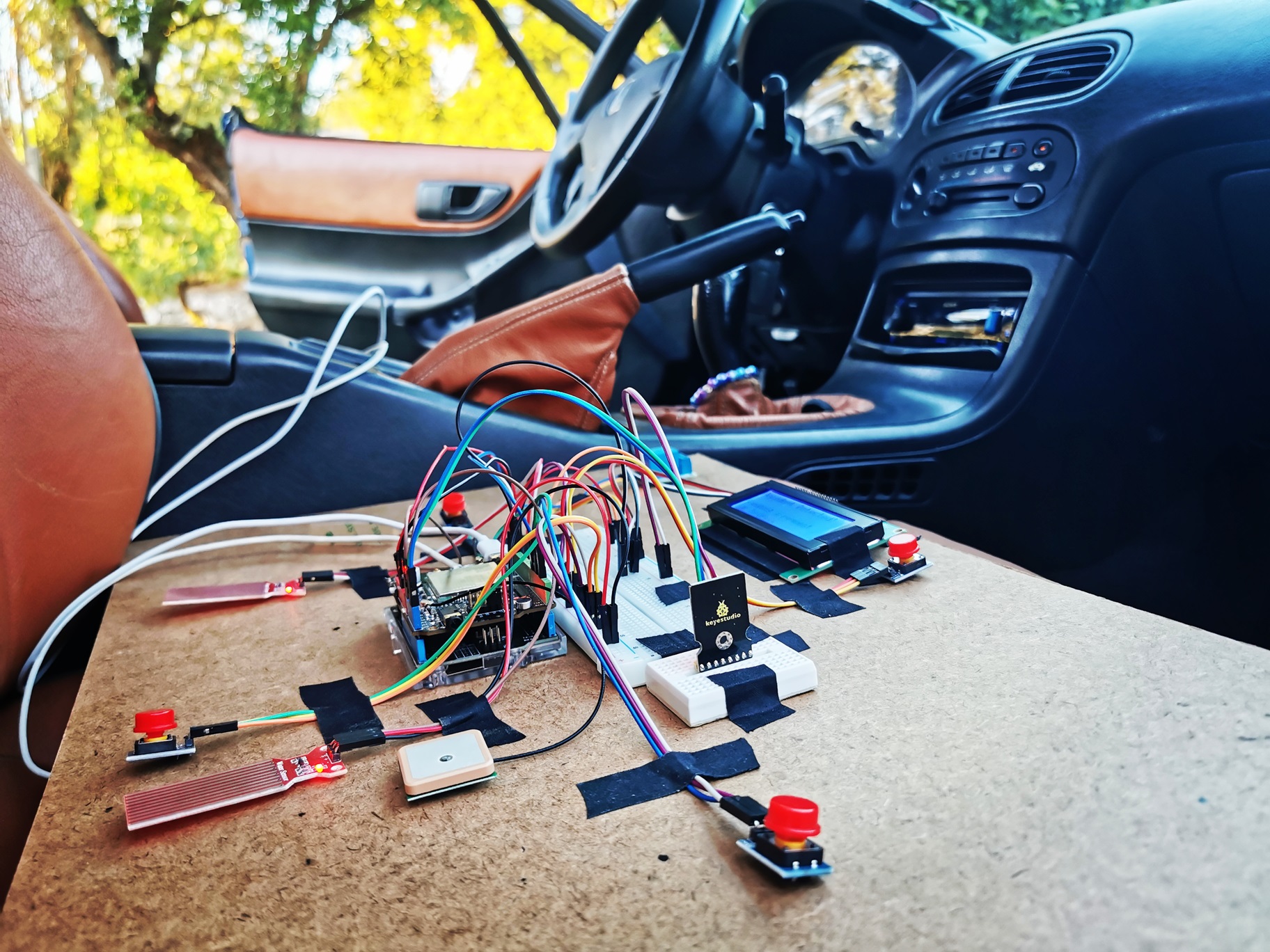}
\caption{Testing prototype on road\label{fig5}.}
\end{figure}

Tests carried out with the "GreenBox" prototype have demonstrated the high effectiveness in accident detection and emergency communication. The system responded promptly to impact, rollover and water detection simulations, sending alert SMS in a matter of seconds. It should be noted that part of the tests were carried out on the road in a vehicle in motion, so that the speed was also obtained, via SMS, and compared with that displayed on the car's speedometer. In Figure~\ref{fig5} we can observe the prototype when testing begins on road.

\begin{figure}[H]
    \centering
    \begin{subfigure}{0.40\textwidth}
        \centering
        \includegraphics[width=\textwidth]{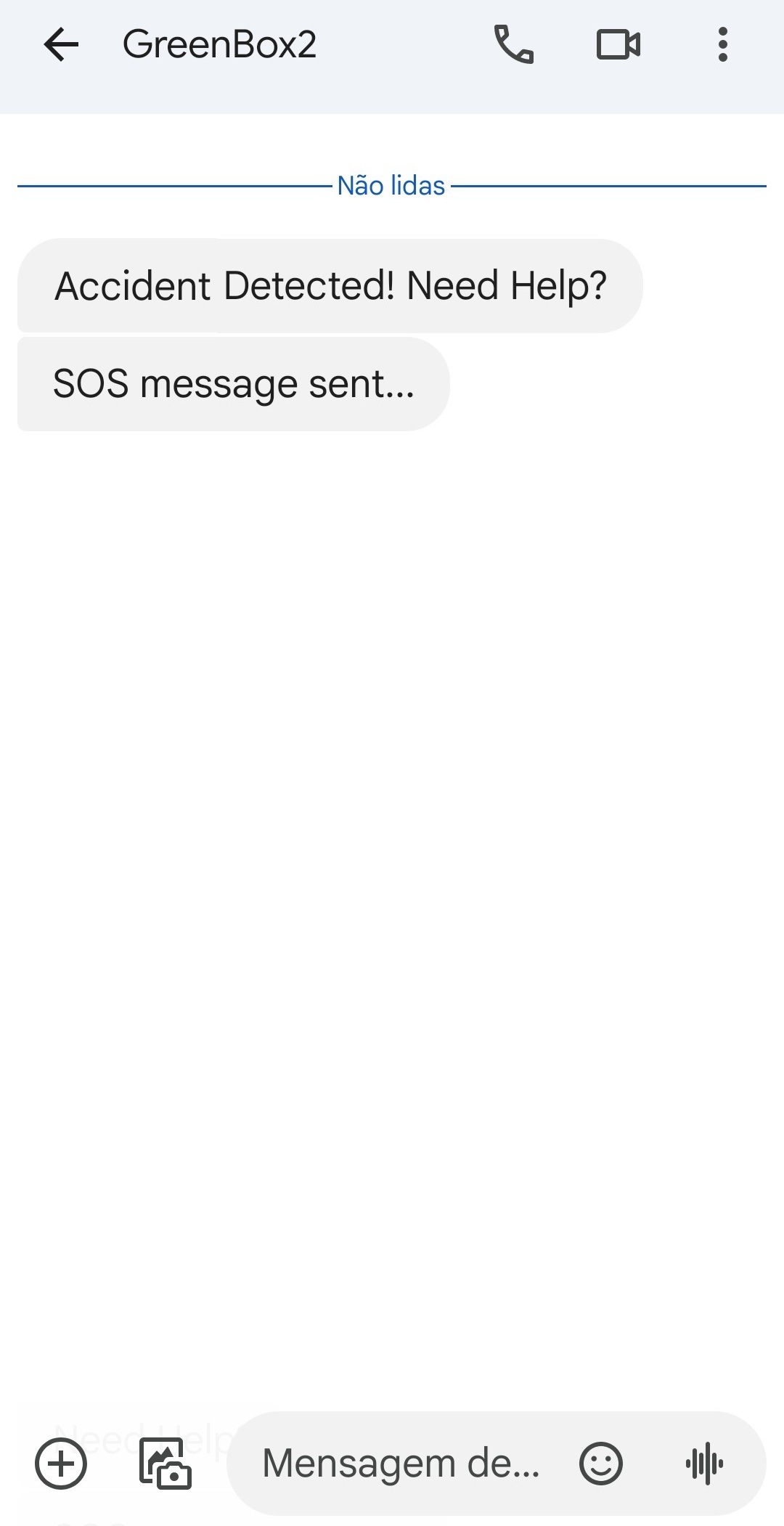}
        \caption{owner/driver sends help SMS to selected recipient}
        \label{fig6}
    \end{subfigure}
    \hspace{0.05\textwidth}
    \begin{subfigure}{0.40\textwidth}
        \centering
        \includegraphics[width=\textwidth]{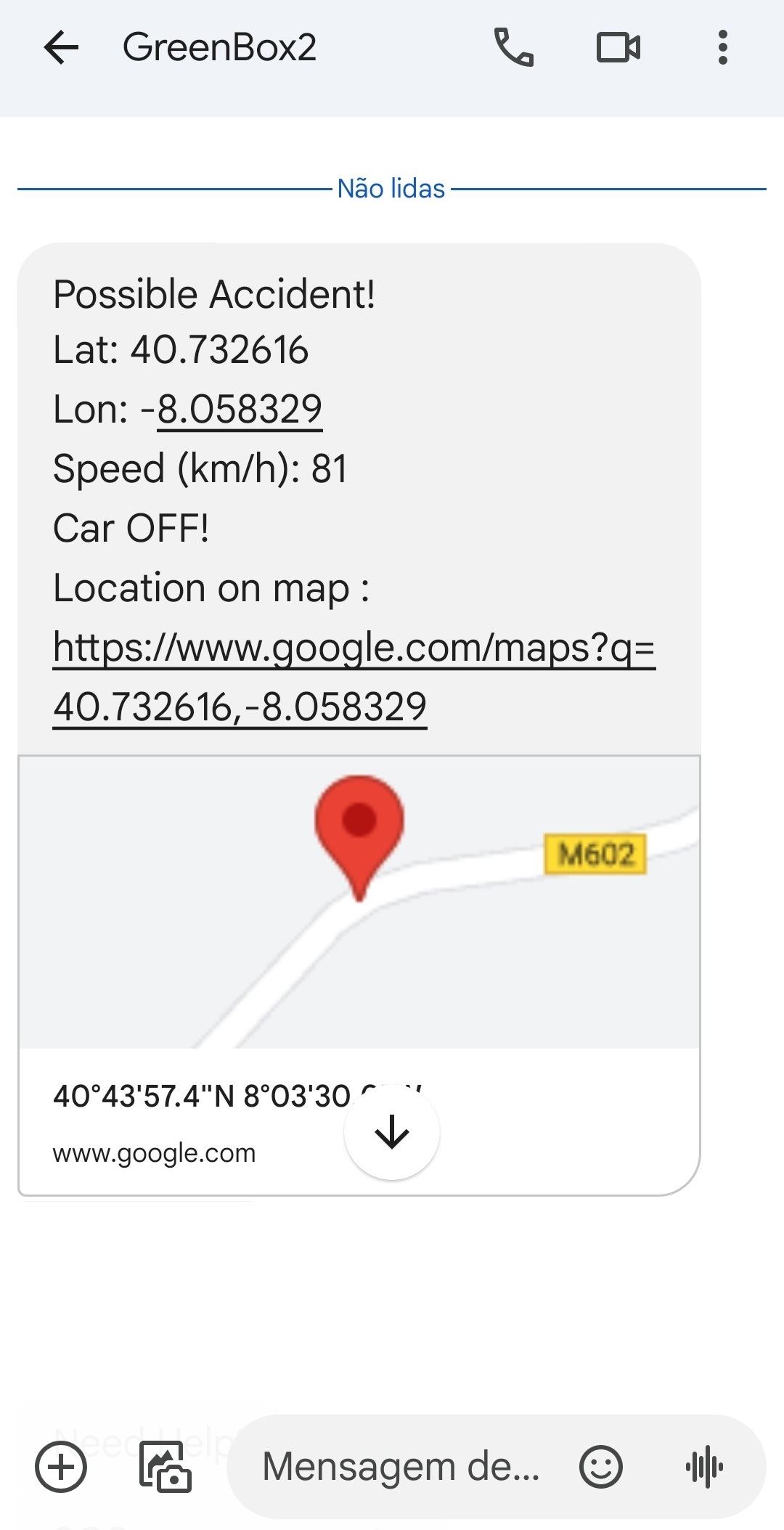}
        \caption{help message received by the selected recipient}
        \label{fig7}
    \end{subfigure}
    \caption{Android SMS Alert Screens}
    \label{fig:android_alerts}
\end{figure}

The accuracy of the GPS location information included in the messages was remarkable, allowing for a quick and accurate location of the vehicle, as well as last known speed, as shown in the Figure~\ref{fig7}.

\begin{figure}[H]
    \centering
    \begin{subfigure}{0.40\textwidth}
        \centering
        \includegraphics[width=\textwidth]{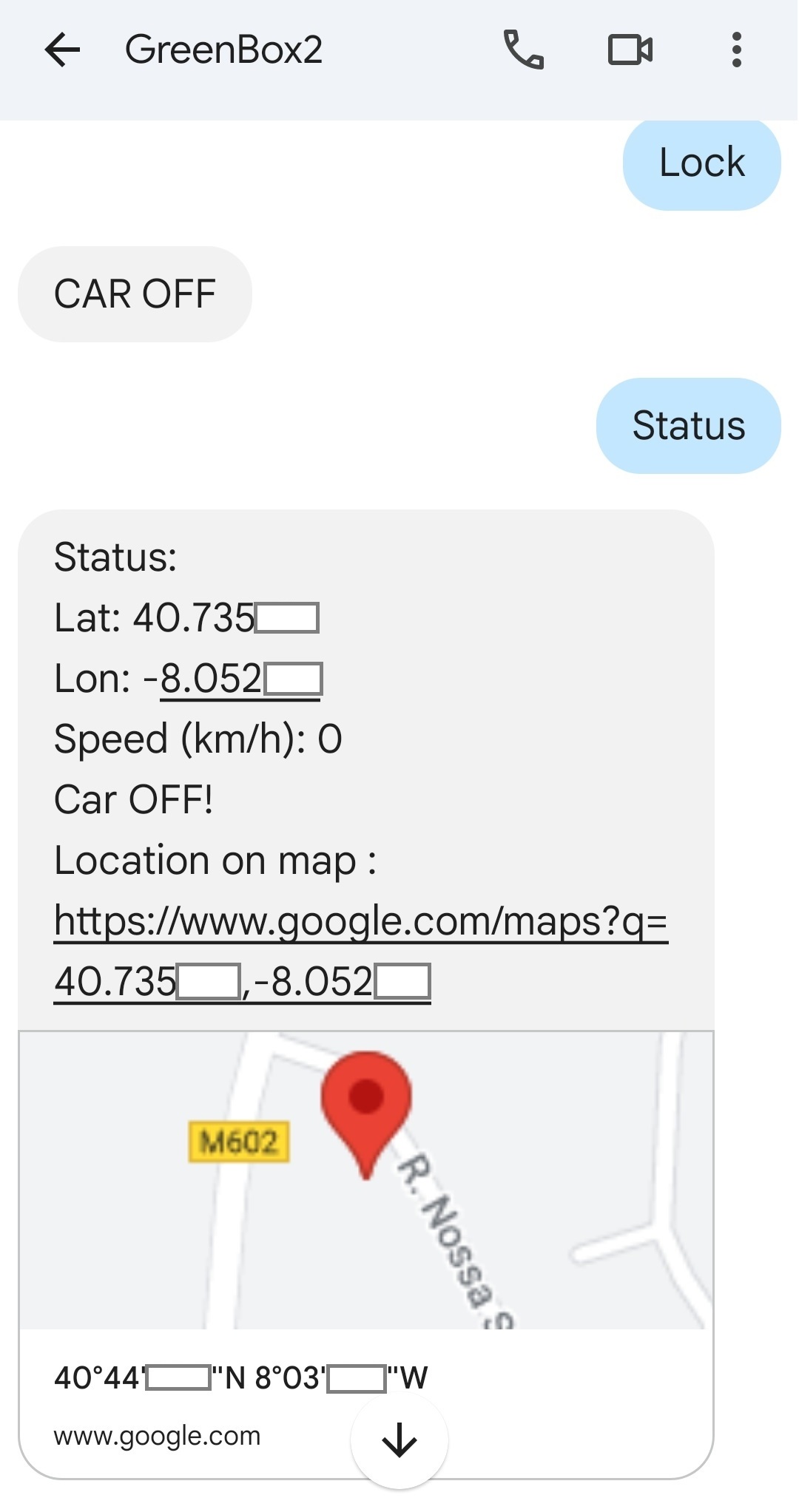}
        \caption{Lock car information}
        \label{fig8}
    \end{subfigure}
    \hspace{0.05\textwidth}
    \begin{subfigure}{0.40\textwidth}
        \centering
        \includegraphics[width=\textwidth]{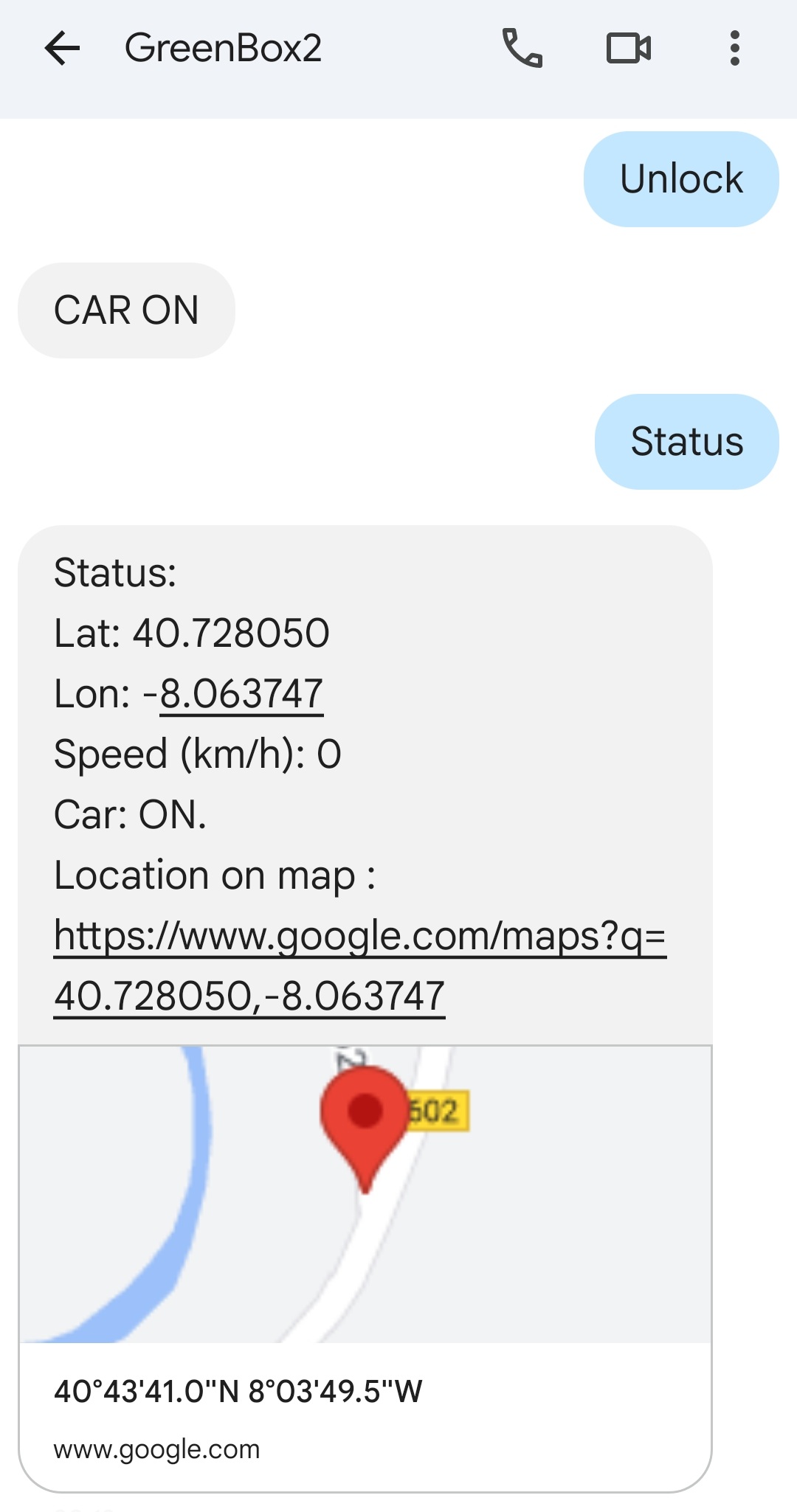}
        \caption{Unlock car information}
        \label{fig9}
    \end{subfigure}
    \caption{Android SMS Response Screens}
    \label{fig:android_response}
\end{figure}

The system also demonstrated excellent responsiveness to remote commands via SMS, such as "status" requests, as also illustrated in Figure~\ref{fig8} and~\ref{fig9}, electronic locking and unlocking of the vehicle's engine, as well as the cancellation of the mechanism when false accident detection, performing these actions quickly and reliably.

The functionality of waiting for confirmation before sending full details of the accident has proven to be a valuable security feature, reducing false alarms. In addition, the implemented SMS sending limit (maximum of 2 messages) proved to be effective in preventing system overload in prolonged emergency situations, for example, when the car remained overturned, that is, when the values, in degrees, of the Pitch and Roll remained outside the parameters established as safe.

\section{Conclusions}

The "GreenBox" system is an integrated road safety solution that uses an Arduino Uno connected to a SIM7600G-H board for accident detection and emergency communication. The system initialises by configuring the pins for sensors (push-buttons, water level sensors, and gyroscope/accelerometer), the relay for motor control, and the GPS module. After start-up, the system enters a continuous loop, where it constantly monitors the impact, tilt, accelerometer/gyroscope, and water level sensors. The GPS updates the vehicle's location and speed every 45 seconds, and this time frame can be adjusted.

When an accident is detected, whether by impact, rollover or flood, the system immediately sends an SMS message to the owner of the system, for example, the driver, asking if he needs help, so that the mechanism can be cancelled, if it is a false positive. If there is no response from the system owner, as the system may be inanimate, within a predefined period, the system automatically sends the details of the accident, including GPS location and the last known speed, to one or more predefined contacts.

The system also responds to SMS commands, allowing you to check the status of the vehicle, such as its location and speed, and to turn the engine on or off remotely. To avoid false positives, the code implements "debounce" techniques on each button, which translates into a millisecond delay in reading the status. In addition, the system limits the number of SMS sent to avoid overloading and manages the vehicle's status (on/off) via a relay connected to the central engine unit, which, as already mentioned, can be controlled remotely.

After several rigorous tests, it is concluded that the "GreenBox" prototype is reliable and can be exploited in order to be commercialized, and its installation will be similar to a traditional alarm system.

The "GreenBox" system is a life-saving solution!

\section*{Acknowledgments}

Authors would like to thank the civil protection entity.

\section*{Conflicts of Interest}

The authors declare no conflict of interest.

\section*{Abbreviations}

The following abbreviations are used in this manuscript:

\begin{center}
\begin{tabular}{@{}ll}
ADAS & Advanced Driver Assistance Systems\\
DNS & Domain Name System\\
GPS & Global Positioning System\\
GSM & Global System for Mobile Communications\\
HTTP & Hypertext Transfer Protocol\\
HTTPS & Hypertext Transfer Protocol Secure\\
IoT & Internet of Things\\
LCD & Liquid Crystal Display\\
ML & Machine Learning\\
SMS & Short Message Service\\
SIM & Subscriber Identity Module\\
TCP/IP & Transmission Control Protocol/Internet Protocol
\end{tabular}
\end{center}

\bibliographystyle{unsrt}
\bibliography{references}

\end{document}